\documentclass[superscriptaddress, twocolumn, pre, longbibliography, nofootinbib]{revtex4-1}
\usepackage{amsmath}
\usepackage{amsbsy}
\usepackage{amssymb}
\usepackage{graphicx}
\usepackage{color}
\usepackage{subfigure}
\usepackage{physics}
\usepackage{soul}
\usepackage{color}
\usepackage{bm}
\usepackage{array}
\usepackage{multirow}
\usepackage{lipsum}
\usepackage[normalem]{ulem}
\usepackage{verbatim}
\usepackage{natbib}
\usepackage{nccmath}
\definecolor{linkcolor}{rgb}{0,0,0.6} 
\usepackage[pdftex,colorlinks=true,
	pdfstartview = FitV,
	linkcolor    = linkcolor,
	citecolor    = linkcolor,
	urlcolor     = linkcolor,	
	hyperindex   = true,
	hyperfigures = false]{hyperref}

\newcommand{\A}{\text{A}}

\newcommand{\bP}{\text{\bf P}}
\newcommand{\T}{T}
\newcommand{\U}{U}
\newcommand{\Pe}{{\rm Pe}}

\newcommand\numberthis{\addtocounter{equation}{1}\tag{\theequation}}

\begin{document} 

\title{From predicting to learning dissipation from pair correlations of active liquids}

\author{Gregory Rassolov*}
\author{Laura Tociu*}
\affiliation{James Franck Institute, University of Chicago, Chicago, IL 60637}
\affiliation{Department of Chemistry, University of Chicago, Chicago, IL 60637}

\author{\'Etienne Fodor}
\affiliation{Department of Physics and Materials Science, University of Luxembourg, L-1511 Luxembourg}

\author{Suriyanarayanan Vaikuntanathan}
\affiliation{James Franck Institute, University of Chicago, Chicago, IL 60637}
\affiliation{Department of Chemistry, University of Chicago, Chicago, IL 60637}
\begin{abstract}
Active systems, which are driven out of equilibrium by local non-conservative forces, can adopt unique behaviors and configurations. An important challenge in the design of novel materials which utilize such properties is to precisely connect the static structure of active systems to the dissipation of energy induced by the local driving. Here, we use tools from liquid-state theories and machine learning to take on this challenge. We first demonstrate analytically for an isotropic active matter system that dissipation and pair correlations are closely related when driving forces behave like an active temperature. We then extend a nonequilibrium mean-field framework for predicting these pair correlations, which unlike most existing approaches is applicable even for strongly interacting particles and far from equilibrium, to predicting dissipation in these systems. Based on this theory, we reveal a robust analytic relation between dissipation and structure which holds even as the system approaches a nonequilibrium phase transition. Finally, we construct a neural network which maps static configurations of particles to their dissipation rate without any prior knowledge of the underlying dynamics. Our results open novel perspectives on the interplay between dissipation and organization out-of-equilibrium.
\end{abstract}

\maketitle
\section{Introduction}
{A}ctive matter is a class of nonequilibrium systems in which energy is consumed at the level of individual components in order to produce autonomous motion~\cite{Marchetti2013, Bechinger2016, Marchetti2018}. These systems exist across many different length- and time-scales, from bacterial swarms~\cite{Elgeti2015} and assemblies of self-propelled colloids~\cite{Palacci2013} to animal groups~\cite{Cavagna2014} and human crowds~\cite{Bartolo2019}. In all these systems, the energy fluxes stemming from individual self-propulsion may lead to complex collective behaviors without any equilibrium equivalent. The possibility of harnessing such behaviors to design materials with novel or useful functions has motivated much research~\cite{Geiss2009} aimed at reliably predicting and controlling the properties of active systems.

Minimal models have been proposed to describe the dynamics of active particles with or without aligning interactions, which can give rise to collective directed motion~\cite{Chate2020, Sood2014} and motility-induced phase separation (MIPS) despite purely repulsive interactions~\cite{Cates2015, Palacci2013} respectively. From these models, the challenge is to establish a nonequilibrium framework, analogous to equilibrium statistical thermodynamics, which connects microscopic details and emergent physics. Some progress has been achieved towards this end by characterizing protocol-based observables, such as pressure~\cite{Brady2014, Solon2015}, surface tension~\cite{Speck2015, Zakine2020}, and chemical potential~\cite{Guioth2019}.

The dissipation induced by microscopic energy fluxes has recently attracted much attention, since it measures the cost to drive the dynamics into nonequilibrium states~\cite{Junco2018, Murrell2019} and to extract work with original protocols~\cite{Pietzonka2019, Liao2020, Ekeh2020}. In particular, it has been shown that dissipation constrains the transport of active particles~\cite{Suri2019, Suri2020}, and that changing dissipation with a dynamical bias changes material properties~\cite{Suri2019}, ultimately inducing phase transitions~\cite{Nemoto2019, Suri2020, GrandPre2020}. Moreover, the dissipation can be connected to some mechanical properties, such as the so-called active pressure, of certain isotropic active matter models~\cite{Solon2015}.

Despite such advancements, understanding how to quantitatively predict the dynamics and structure of many-body active systems remains largely an open challenge~\cite{Gompper_2020}. Many of the theoretical approaches used to predict the structure of active matter rely on either equilibrium mappings~\cite{Szamel2015, Rein2016, Wittmann2017b, Szamel2019} or weak-interaction approximations~\cite{Suri2019, Suri2020}, thus limiting their applicability.

In this work (along with its companion paper in~\cite{Tociu2022}), we use tools from liquid-state theories and machine learning to take on these challenges. We develop a mean-field theory which quantitatively connects the rate of energy dissipation to static two-point density correlations and whose applicability and ease of implementation surpasses existing approaches. Our results provide a gateway towards controlling the structure and properties of a nonequilibrium many-body system through tuning the dissipation rate, and they also demonstrate how machine learning may be harnessed to guide such a process. For instance, there is a close connection between active pressure and dissipation for certain models of active matter~\cite{Solon2015}, and here we express dissipation purely in terms of the isotropic two-body configurational correlation function. This suggests that we can anticipate how the pressure of a non-equilibrium active liquid is directly related to isotropic two-body correlations. Such a \emph{virial}-like connection may provide intuition for how the structure of a non-equilibrium fluid may change as external force conditions are modulated.

The paper is organized as follows. First, in Section~\ref{model}, we describe the model of an active liquid that we use in this work, which is an assembly of isotropic self-propelled particles. Second, in Section~\ref{smalltau}, we derive analytical expressions to argue that dissipation and pair correlations are closely related for these particles in \emph{at least} the regime of driving forces effectively acting as additional thermal noise. Third, motivated by this conclusion, in Section~\ref{meanfield} we extend a nonequilibrium mean-field theory that we first introduced in~\cite{Tociu2022} for predicting pair correlations out of equilibrium \textendash~even for strongly interacting particles \textendash~to evaluate energy dissipation. Unlike other effective representations of active dynamics~\cite{Szamel2015, Rein2016, Wittmann2017, Szamel2019} and our derivation in Section~\ref{smalltau}, we do \emph{not} rely on any equilibrium approximation in this mean-field theory, thus allowing all nonequilibrium features to be retained, and in particular for the dissipation to be properly evaluated. Combining our results, in Section~\ref{simulations} we then elucidate the relationship between static density correlations and dissipation over various regimes of activity and interaction strength. Taken together, our results exploring this relation suggest that dissipation can be reliably predicted without analyzing properties of the system other than isotropic two-body correlations. In contrast, existing approaches to quantify dissipation rely on evaluating either the violation of fluctuation-response relations~\cite{Sasa2005, Toyabe2010, Ahmed2016, Nardini2017, Mizuno2018}, currents~\cite{Barato2015, Gingrich2016, Gladrow2016, Li2019}, or irreversibility of trajectories~\cite{Roldan2018, Parrondo2019}. Lastly, as a proof of principle, in Section~\ref{neuralnet} we put forward a machine learning architecture, trained with static configurations, which accurately infers the dissipation rate without any information about the underlying dynamics.


\section{Model and overview}
\label{model}

\subsection{Details of model active matter system}

We consider a conventional model of active matter consisting of $N$ interacting self-propelled particles, often referred to as Active Ornstein Uhlenbeck Particles~\cite{Szamel2014, Maggi2015, Nardini2016}, with two-dimensional overdamped dynamics:
\begin{equation}\label{eq:EOM}
	\dot{\bf r}_i = -\frac{1}{\gamma}\nabla_i \sum_{j\neq i} U({\bf r}_i-{\bf r}_j) + \frac{{\bf f}_i}{\gamma} + {\boldsymbol\xi}_i ,
\end{equation}
where $U$ is the pair-wise potential, $\gamma$ is a friction coefficient, and $i,j$ are particle indices. The terms $\{{\boldsymbol\xi}_i,{\bf f}_i\}$ embody, respectively, the thermal noise and the self-propulsion velocity. They have Gaussian statistics with zero mean and uncorrelated variances, given by:
\begin{align}\label{eq:EOM_noise}
    & \langle\xi_{i\alpha}(t)\xi_{j\beta}(0)\rangle = \frac{2T}{\gamma} \delta_{ij}\delta_{\alpha\beta}\delta(t) \nonumber , \\
    & \langle f_{i\alpha}(t) f_{j\beta}(0) \rangle = \frac{\gamma T_{\rm A}}{\tau} \delta_{ij} \delta_{\alpha\beta} e^{-|t|/\tau} ,
\end{align}
where $\tau$ is the persistence time, $T$ is thermal temperature, and $\{\alpha,\beta\}$ index spatial dimensions. For a vanishingly small persistence ($\tau \rightarrow 0$), the system reduces to a set of passive Brownian particles at temperature $T+T_{\rm A}$. At sufficiently high $T_{\rm A}$ and $\tau$, the system undergoes phase separation even with a purely repulsive interparticle potential $U$~\cite{Nardini2016}. All details pertaining to the simulations, run in two dimensions for all of what follows, can be found in Materials and Methods.

Following standard definitions of stochastic thermodynamics~\cite{Sekimoto1998, Seifert2012}, the dissipation is defined as the rate of work done by the thermostat on particles: ${\cal J} = \gamma \sum_i \langle \dot{\bf r}_i\cdot(\dot{\bf r}_i - {\boldsymbol\xi}_i) \rangle$. It can be decomposed into a constant term, independent of interactions, and a contribution from interactions between particles modulated by the driving forces: ${\cal J} = N (T_{\rm A}/\tau - \dot w)$. Here we have introduced the \emph{active work} per particle~\cite{Suri2019, Suri2020, Junco2018}:
\begin{equation}\label{eq:dotW_microscopic}
    \dot{w} = \frac{1}{N\gamma} \sum_{j \neq i} \langle {\bf f}_i \cdot \nabla_i U \left({\bf r}_i - {\bf r}_j\right) \rangle .
\end{equation}
Hereafter, we use the term dissipation only to refer to this active work per particle contribution to the total dissipation. It is large when driving forces push against conservative repulsive interparticle forces and hold particles in close proximity to each other. The expression~\eqref{eq:dotW_microscopic} illustrates that, for a generic many-body active system, any particle-based evaluation of dissipation requires measuring the local polarization ${\bf f}_i$. This is notoriously difficult in experimental systems of isotropic active particles, for which the local driving direction is not encoded in the particle shape. Additionally, we note that~\eqref{eq:dotW_microscopic} is closely related to the density-dependent average speed of the particles~\cite{Solon2015}, which is directly connected to the active swim pressure in such systems~\cite{Brady2014, Solon2015}.

\subsection{Review: Exactly relating dissipation and spatial correlations}

Following~\cite{Suri2019}, by using It\^o calculus and assuming that the system has reached a steady state with no change in average total $U$, we can derive an exact relationship between $\dot{w}$ and two- and three-body correlations for AOUPs:
\begin{equation}\label{eq:dotW_Ito}
\begin{aligned}
    \dot{w}  =  & \frac{\rho_0}{\gamma} \int d\mathbf{r}  g\left(\mathbf{r}\right)  \left[  \left( \nabla U\left(\mathbf{r}\right) \right)^2  -  T \nabla^2 U\left(\mathbf{r}\right)  \right]   \\
    & + \frac{\rho_0^2}{\gamma} \iint d\mathbf{r} d\mathbf{r'}  g_3\left(\mathbf{r,r'}\right)  \left(  \nabla U\left(\mathbf{r}\right)) \cdot (\nabla U\left(\mathbf{r'}\right)  \right) ,
\end{aligned}
\end{equation}
with the correlation functions defined as:
\begin{equation}
\begin{aligned}\label{eq:g2g3}
    g & \left(\mathbf{r}\right)  =  \frac{1}{N \rho_0}  \sum_{j \neq i}  \langle  \delta \left( \mathbf{r} - {\bf r}_i + {\bf r}_j \right)  \rangle, \\
    g_3 & \left(\mathbf{r,r'}\right) = \frac{1}{N \rho_0^2}  \sum_{k \neq j \neq i}  \langle  \delta \left( \mathbf{r} - {\bf r}_i + {\bf r}_j\right)  \delta \left( \mathbf{r'} -{\bf r}_i + {\bf r}_k\right)  \rangle,
\end{aligned}
\end{equation}
where $\rho_0$ is the average particle density. We emphasize that this is an exact relation for AOUPs at steady state, true for any $\rho_0$ or $T_{\rm A}$. With a vanishing rate of work, this expression recovers the first order of the equilibrium Yvon-Born-Green (YBG) hierarchy for two-component fluids~\cite{Hansen}. Note that in the limit of small $\rho_0$, $\dot{w}$ will unsurprisingly be related only to pair correlations in the system, as contributions from the three-body correlation term are suppressed.

\section{Results}
\label{Results}

\subsection{Connection between dissipation and pair correlations for short persistence times}
\label{smalltau}

We now aim to relate dissipation to pair correlations beyond the simple low-density limit mentioned above. In this section, we accomplish this in the limit $\tau \rightarrow 0$, wherein the properties of the non-equilibrium system can be well defined in terms of an effective temperature. In particular, we assume $\tau$ is sufficiently small that a nonequilibrium system at thermal temperature $T$ and with activity set by $T_{\rm A}$ has the same spatial correlations as an equilibrium system at thermal temperature $T + T_{\rm A}$~\cite{Suri2020}:
\begin{equation}\label{eq:app_1}
    g_{\rm neq}(\mathbf{r};T) \underset{\tau\to0}{\longrightarrow} g_{\rm eq}(\mathbf{r};T+T_{\rm A}).
\end{equation}
From~\eqref{eq:dotW_Ito}, an equilibrium system with zero dissipation at thermal temperature $T + T_{\rm A}$ obeys the following YBG energy balance relation:
\begin{equation}\label{eq:app_2}
\begin{aligned}
    &-\rho_0 \iint d\mathbf{r} d\mathbf{r'}  g_{3,\rm eq}\left(\mathbf{r,r'};T+T_{\rm A}\right)  (  \nabla U\left(\mathbf{r}\right)) \cdot (\nabla U\left(\mathbf{r'}\right) )  \\
    &= \int d\mathbf{r}  g_{\rm eq}\left(\mathbf{r};T+T_{\rm A}\right)  \left[  \left( \nabla U\left(\mathbf{r}\right) \right)^2  -  \left(T+T_{\rm A}\right) \nabla^2 U\left(\mathbf{r}\right)  \right] ,
\end{aligned}
\end{equation}
while the corresponding non-equilibrium system at thermal temperature $T$ satisfies the following:
\begin{equation}\label{eq:app_3}
\begin{aligned}
    \dot{w} &= \frac{\rho_0^2}{\gamma} \iint d\mathbf{r} d\mathbf{r'}  g_{3,\rm neq}\left(\mathbf{r,r'};T\right)  (  \nabla U\left(\mathbf{r}\right)) \cdot (\nabla U\left(\mathbf{r'}\right))  \\
    &\quad+ \frac{\rho_0}{\gamma} \int d\mathbf{r}  g_{\rm neq}\left(\mathbf{r};T\right)  \left[  \left( \nabla U\left(\mathbf{r}\right) \right)^2  -  T \nabla^2 U\left(\mathbf{r}\right)  \right] .
\end{aligned}
\end{equation}
Combining Eqs.~(\ref{eq:app_1}-\ref{eq:app_3}), we obtain the following expression for $\dot{w}$:
\begin{equation}\label{eq:dotW_smalltau}
    \dot{w} \underset{\tau\to0}{\longrightarrow} -\frac{\rho_0 T_{\rm A}}{\gamma} \int d\mathbf{k}  k^2 g_{\rm eq}\left(\mathbf{k};T+T_{\rm A}\right) U\left(\mathbf{-k}\right) .
\end{equation}
While this straightforwardly relates dissipation and pair correlations in the small $\tau$ regime, in~\cite{Suri2019} a robust empirical relationship between $\dot{w}$ and the \emph{difference} in pair correlations between the system away from and at equilibrium, $g_{\rm neq}\left(\mathbf{r}; T\right) - g_{\rm eq}\left(\mathbf{r}; T\right) \equiv \Delta g\left(\mathbf{r}\right)$, was observed to hold even in many regimes beyond the small $\tau$ regime:
\begin{equation}\label{eq:dotW_tildeI}
    \dot{w} \propto \frac{\rho_0}{\gamma} \int d\mathbf{r}  \Delta g\left(\mathbf{r}\right)  \left[  \left( \nabla U\left(\mathbf{r}\right) \right)^2  -  T \nabla^2 U\left(\mathbf{r}\right)  \right]  \equiv  \tilde{I} .
\end{equation}
$\tilde{I}$, defined by this expression, is the main form in which we consider pair correlations in the remainder of this work. For repulsive AOUPs, $g(\mathbf{r})$ at small $\mathbf{r}$ increases as driving forces are increased, pushing particles into each and leading to enhanced clustering; this is reflected by larger values of $\tilde{I}$. The constant of proportionality is dependent on $\rho_0$, $U$, and $\tau$, but independent of the P\'eclet number $\Pe \equiv (\sigma/T) \sqrt{T_{\rm A}/\tau}$, which compares the relative strength of the active drive with respect to thermal fluctuations, and in which $\sigma$ denotes the typical interaction range (taken to be unity in what follows). We therefore also seek to build on~\eqref{eq:dotW_smalltau} to analytically justify the empirical relationship~\eqref{eq:dotW_tildeI} in at least the small $\tau$ regime. To do this, we simply subtract~\eqref{eq:dotW_Ito} for a system at thermal temperature $T$ with $T_{\rm A}=0, \dot{w}=0$ from the same expression for a system at thermal temperature $T$ with $T_{\rm A}>0, \dot{w}>0$, yielding:
\begin{equation}\label{eq:dotW_tildeI_exact}
\begin{aligned}
    \dot{w} = & ~\frac{\rho_0}{\gamma} \int d\mathbf{r}  \Delta g\left(\mathbf{r}\right)  \left[  \left( \nabla U\left(\mathbf{r}\right) \right)^2  -  T \nabla^2 U\left(\mathbf{r}\right)  \right]  \\
    & + \frac{\rho_0^2}{\gamma} \iint d\mathbf{r} d\mathbf{r'}  \Delta g_3\left(\mathbf{r,r'}\right)  ( \nabla U\left(\mathbf{r}\right)) \cdot (\nabla U\left(\mathbf{r'}\right) ) \\
    =& ~\tilde{I} + \frac{\rho_0^2}{\gamma} \iint d\mathbf{r} d\mathbf{r'}  \Delta g_3\left(\mathbf{r,r'}\right)  (\nabla U\left(\mathbf{r}\right)) \cdot (\nabla U\left(\mathbf{r'}\right) ).
\end{aligned}
\end{equation}
where we have used the definition of $\tilde I$ in Eq.~\eqref{eq:dotW_tildeI}. Note that unlike~\eqref{eq:dotW_smalltau}, the relation~\eqref{eq:dotW_tildeI_exact} is exact and does not yet assume small $\tau$.

\begin{figure}
    \centering
    \includegraphics[scale=0.1, clip=True]{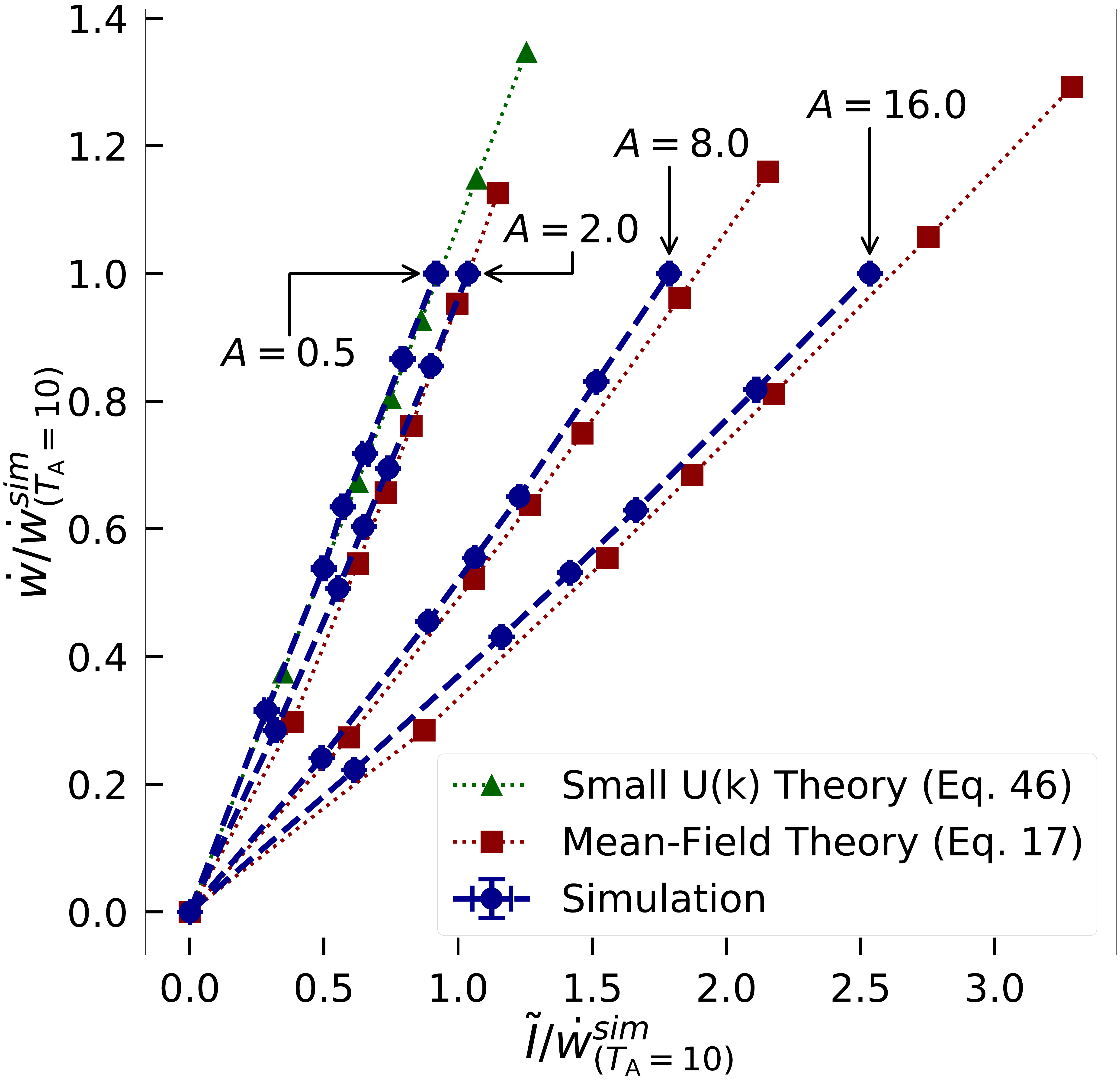}
    \caption{
	Connecting dissipation and structure for harmonic AOUPs.
	Theory and simulation results for the dissipation $\dot{w}$ and the pair correlation observable $\tilde I$ yield a linear relation $\dot w = \alpha \tilde I$ for which the slope $\alpha$ varies with the interaction strength $A$ (labeled). For the weakest interactions shown ($A=0.5$), the slope predicted by our mean-field theory expression for weakly interacting particles described in Appendix~\ref{sec:appB} (green, triangles;~\eqref{eq:dotW_MF_final} for $\dot{w}$ and~\eqref{eq:dotW_tildeI_Gaussian} for $\tilde{I}$ relative to $\dot{w}$) and that measured numerically (blue, circles) agree very well. For stronger interactions ($A>0.5$), our final and more general version of the mean-field theory with the $c_{\rm eq}({\bf k})$ substitution (red, squares;~\eqref{eq:dotW_MF_c} for $\dot{w}$ and predictions from~\eqref{eq:hk_MF_c} inserted into~\eqref{eq:dotW_tildeI} for $\tilde{I}$) leads to predicted slopes that are close to those measured from simulations (blue, circles). These results demonstrate that the dissipation-structure relation is robust across different values of $A$ and that our mean-field theory accurately captures it.
	Parameters: harmonic interparticle potential, $\rho_0=1$, $\tau=1$, $T_{\rm A}\in[0,10]$, $T=1$, $\gamma=1$. Note that all values for a given $A$ are rescaled by the corresponding value of $\dot{w}$ at $T_{\rm A}=10$ from simulations.
	}
    \label{Fig:wvsI_harmonic}
\end{figure}

To further simplify this relation, we assume that we are in a regime such that density fluctuations are Gaussian and small relative to the average density $\rho_0$, which is most reasonable when particles are weakly interacting. This allows us to rewrite $\Delta g_3$ as the sum of three $\Delta g$ terms, two of which go to 0 when integrated over $\mathbf{r}$ and $\mathbf{r'}$ (see Appendix~\ref{sec:appA}), yielding in real and Fourier space:
\begin{equation}\label{eq:dotW_tildeI_Gaussian}
\begin{aligned}
    \dot{w} & = ~\tilde{I} + \frac{\rho_0^2}{\gamma} \iint d\mathbf{r} d\mathbf{r'}  \Delta g\left(\mathbf{r,r'}\right)  ( \nabla U\left(\mathbf{r}\right)) \cdot (\nabla U\left(\mathbf{r'}\right) ) \\
    & = \tilde{I} + \frac{\rho_0^2}{\gamma} \int d\mathbf{k} k^2 \Delta g\left(\mathbf{k}\right) U\left(\mathbf{k}\right) U\left(\mathbf{-k}\right) .
\end{aligned}
\end{equation}
From~\eqref{eq:dotW_tildeI_Gaussian} and~\eqref{eq:dotW_smalltau}, we then get
\begin{equation}
    \begin{aligned}
    \tilde I \underset{\tau\to0}{\longrightarrow} & -\frac{\rho_0}{\gamma} \int d\mathbf{k}  k^2 g_{\rm eq}\left(\mathbf{k};T+T_{\rm A}\right) U\left(\mathbf{-k}\right) \Big[ \rho_0 U\left(\mathbf{k}\right) + T_{\rm A} \Big]
    \\
    &+ \frac{\rho_0^2}{\gamma} \int d\mathbf{k} k^2 g_{\rm eq}\left(\mathbf{k};T\right) U\left(\mathbf{k}\right) U\left(\mathbf{-k}\right) .
    \end{aligned}
\end{equation}
We can use the Ornstein-Zernike relation for equilibrium liquids~\cite{Hansen}, which relates the pair correlation function $g_{\rm eq}$ with the direct correlation function $c_{\rm eq}$ as
\begin{equation}\label{eq:c}
g_{\rm eq}\left(\mathbf{k};T\right) = \frac{c_{\rm eq}\left(\mathbf{k};T\right)}{1-\rho_0 c_{\rm eq}\left(\mathbf{k};T\right)} .
\end{equation}
In the limit of weak interparticle interactions, $c_{\rm eq}(\mathbf{k};T) \approx -U(\mathbf{k})/T$~\cite{Hansen}, yielding
\begin{equation}
\label{eq:dotW_smalltau_smallU}
\begin{aligned}
    \dot{w} &\underset{\tau\to0}{\longrightarrow} \frac{\rho_0 T_{\rm A}}{\gamma} \int d\mathbf{k}  k^2 \frac{U\left(\mathbf{k}\right)}{T+T_{\rm A}+\rho_0 U\left(\mathbf{k}\right)} U\left(\mathbf{-k}\right) ,
\\
    \tilde{I} &\underset{\tau\to0}{\longrightarrow} \frac{\rho_0 T_{\rm A}}{\gamma} \int d\mathbf{k} \frac{k^2 U\left(\mathbf{k}\right) U\left(\mathbf{-k}\right)}{T+T_{\rm A}+\rho_0 U\left(\mathbf{k}\right)} \frac{T}{T+\rho_0 U\left(\mathbf{k}\right)} .
\end{aligned}
\end{equation}
If $\rho_0 U\left(\mathbf{k}\right)$ is small compared to $T$, and if the integrand in~\eqref{eq:dotW_smalltau_smallU} has a single sharp peak at wavenumber $\mathbf{k_c}$, then these results taken together suggest the following:
\begin{equation}\label{eq:dotW_tildeI_simple}
    \dot{w} = \frac{T+\rho_0 U\left(\mathbf{k_c}\right)}{T} \tilde{I} .
\end{equation}
This relationship does not depend on the strength of driving forces in the system, i.e. it is independent of $T_{\rm A}/\tau$, and leads us to expect that $\dot{w}>\tilde{I}$ in systems satisfying the assumptions listed above. Empirically, we observe this inequality to hold for weakly interacting particles, while the opposite holds if interactions are strong; see Fig.~\ref{Fig:wvsI_harmonic}. Additionally, note that in the further limit of small $\rho_0$ or \emph{very} weak $U(\mathbf{k})$ relative to $T$,~\eqref{eq:dotW_tildeI_simple} simplifies to $\dot{w} = \tilde{I}$.

\subsection{Mean-field theory for dissipation of AOUPs}
\label{meanfield}

The expressions~\eqref{eq:dotW_smalltau_smallU} are derived in the limit of small $\tau$. However, as mentioned previously, in~\cite{Suri2019} it was found that $\dot{w} \propto \tilde{I}$ was empirically observed to hold across many regimes. This includes regimes with $\tau \gg 0$, for which non-equilibrium activity behaves qualitatively very differently from an enhancement to thermal temperature. To gain additional insight into why dissipation can be related to just pair correlations with no reference to three-body correlations in such systems, we now turn to a mean-field theory which makes no assumptions about $\tau$ that, in our companion paper~\cite{Tociu2022}, we apply to predicting pair correlations for out-of-equilibrium AOUP systems.

The mean field theory is outlined in Appendix~\ref{sec:appB}. In short, we consider a single arbitrary AOUP as a tracer particle and the surrounding AOUPs as a density field with linearizable dynamics; this assumes a regime of weak interactions and high densities. Expressing $\dot{w}$ and $h(\mathbf{k})$ in terms of the density field dynamics, keeping the lowest-order terms, and simplifying yields mean-field expressions~\eqref{eq:dotW_MF_final} and~\eqref{eq:hk_MF} suitable for weakly interacting particles. The final expressions for $\dot{w}$ and $h(\mathbf{k})$, applicable to strongly interacting particles, are obtained by using $-Tc_{\rm eq}(\mathbf{k})$ in place of $U(\mathbf{k})$ to account for higher-order effects. These expressions are

\begin{equation}\label{eq:dotW_MF_c}
\begin{aligned}
    \dot{w} &= \frac{\rho_0 T_{\rm A}}{\gamma} \int \frac{d{\bf k}}{(2\pi)^d} {\bf k}^4 T c_{\rm eq}({\bf k}) U({\bf k}) \dfrac{ \hat{G}({\bf k}) + T}{\hat{G}({\bf k})} \\
    &\quad\times \int_{-\infty}^0 ds e^{{\bf k}^2 \left((\hat{G}({\bf k}) + T)s + R(s)\right)}(1-e^{s/\tau}) ,
\end{aligned}
\end{equation}
and
\begin{equation}\label{eq:hk_MF_c}
    h({\bf k}) = {\bf k}^2 T c_{\rm eq}({\bf k}) \dfrac{ \hat{G}({\bf k}) + T}{\hat{G}({\bf k})} \int_{-\infty}^0 ds e^{{\bf k}^2 \left( (\hat{G}({\bf k}) + T)s + R(s)\right)} ,
\end{equation}
where $\hat{G}(\mathbf{k})=T\left(1-\rho_0 c_{\rm eq}(\mathbf{k})\right)$ and $R(s) \equiv T_{\rm A} s + T_{\rm A} \tau ( 1- e^{s/\tau})$. In this work, we primarily use~\eqref{eq:hk_MF_c} to obtain predictions of $\tilde{I}$ by converting $h(\mathbf{k})$ to real space and inserting estimates of $g_{\rm neq}(\mathbf{r})$ into the original definition of $\tilde{I}$ in~\eqref{eq:dotW_tildeI}.

In the following section, we demonstrate that~\eqref{eq:dotW_MF_c} and~\eqref{eq:hk_MF_c} yield predictions for particles without hard repulsive cores that are consistent with measured results from simulations in predicting the relationship between $\dot{w}$ and $\tilde{I}$. For hard-core repulsive AOUPs, we have found that even small errors in predictions at large $\mathbf{k}$ make the mean-field theory less usable for obtaining quantitative predictions of $\tilde{I}$. However, the simulation results themselves suggest that a strong relationship between $\dot{w}$ and $\tilde{I}$ still exists. In the subsequent section we demonstrate that a neural network is able to learn it accurately when provided with only static information about interparticle distances, buttressing the evidence that this relationship is robust.

\begin{figure*}
    \centering
    \includegraphics[scale=0.096, clip=True]{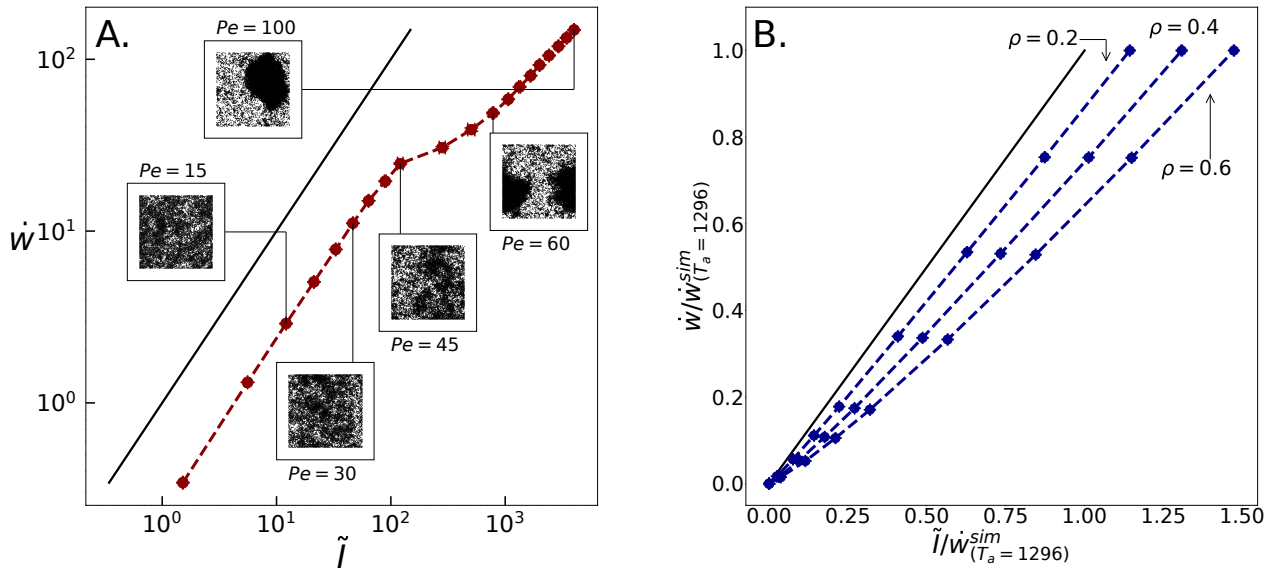}
    \caption{
	Connecting dissipation and structure for strongly interacting AOUPs.
	(a)~The linear relation between $\dot w$ and $\tilde{I}$, as measured numerically, persists up to values of $\Pe$ that are within 25\% of that at which MIPS occurs quickly and spontaneously. The insets show snapshots of the system at different values of $\Pe$. The black solid line is a guideline corresponding to $\dot w=\tilde I$.
	Parameters: WCA potential, $\rho_0=0.75$, $\tau=100/3$, $T=1$, $\gamma=100$, $\Pe \in[5,100]$, and $\Pe \equiv (\sigma/T) \sqrt{T_{\rm A}/\tau}$ with $\sigma=1$.
	Note that the plot is logarithmic, so in the region where simulation data is parallel to the guideline, $\dot w = \alpha \tilde I$ holds.
	(b)~Simulation results for $\dot{w}$ vs $\tilde I$ for particles with the Yukawa interparticle potential (dashed blue lines) at various densities (labeled). The ratio between $\dot{w}$ and $\tilde I$ is essentially constant for values of $T_{\rm{A}}$ ranging from weak to moderately strong driving, with a decreasing value of the slope as $\rho$ increases. The solid black line corresponds to a slope of 1, indicating $\dot{w}=\tilde{I}$. Parameters: Yukawa potential, $A=50$, $\kappa=4.0$, $\tau=1$, $T=1$, $\gamma=1$, $T_{\rm{A}}\in[0,1296]$.
	}
    \label{Fig:wvsI_strong}
\end{figure*}

\subsection{Numerical results}
\label{simulations}

The empirical connection in~\eqref{eq:dotW_tildeI} was reported in~\cite{Suri2019} for particles interacting through the Weeks-Chandler-Anderson (WCA) potential. We now further numerically explore the robustness of this connection in different regimes and demonstrate that the mean-field results in~\eqref{eq:dotW_MF_c} and~\eqref{eq:hk_MF_c} provide it with an analytical justification.

We simulate particles interacting via the short-ranged harmonic potential, given by $U({\bf r})=A (1-r)^2$ for $r<1$, at multiple values of $T_{\rm A}$ and $A$. In Fig.~\ref{Fig:wvsI_harmonic}, we plot the estimates of $\dot{w}$ and $\tilde I$ obtained from simulations and from our analytical mean-field theory. For every $A$, the simulations are performed at values of $T_{\rm A}$ within the range $[0,10]$, with fixed persistence time $\tau=1$ and density $\rho_0=1$. The simulation results show an essentially constant ratio between $\dot{w}$ and $\tilde I$ for all values of $A$ (blue lines with circles). For the most weakly interacting AOUPs shown with $A=0.5T$, our analytical predictions (green line with triangles pointing down) are made using our original mean-field theory for dissipation of weakly interacting particles~\eqref{eq:dotW_MF_final} and the compact expression from~\eqref{eq:dotW_tildeI_Gaussian} which assumes Gaussian density fluctuations. For higher values of $A$ (red lines with triangles pointing up), we use~\eqref{eq:dotW_MF_c} for $\dot{w}$ and numerically integrate the expression defining $\tilde{I}$ in~\eqref{eq:dotW_tildeI} by substituting $h$ as given in~\eqref{eq:hk_MF_c}, using $c_{\rm{eq}}$ measured in simulations as input.

In both cases, the analytical predictions for $\dot{w}$ and $\tilde I$ deviate moderately from the simulation data. As discussed in~\cite{Tociu2022}, given the errors in the predicted structure $h$ at short wavelengths (reflecting changes in the structure of the system on large length scales as activity is increased), these deviations are not very surprising. Crucially, despite the deviations in absolute values, both of our predictions are able to accurately reproduce the measured value of the ratio between $\dot{w}$ and $\tilde{I}$ for each value of $A$. Additionally, we note that for both $A=0.5T$ and $A=2T$, $\dot{w}>\tilde{I}$ holds, as was qualitatively suggested by~\eqref{eq:dotW_smalltau_smallU} for weakly interacting particles in the limit of small $\tau$ (although here the system is not close to this limit). Overall, these results illustrate that our mean-field theory, designed to capture nonequilibrium liquid properties, accurately predicts the dissipation-structure relation inherent to active liquids, and that the use of $-Tc_{\rm eq}({\bf k})$ allows it to be extended to strongly interacting particles.

The linear connection between $\dot{w}$ and $\tilde I$ eventually fails at larger values of $T_{\rm A}$ for the harmonic potential (Fig.~\ref{Fig:A1}). We anticipate that this stems from the fact that such a potential allows for a complete overlap of particles at sufficiently strong driving. For liquids with a hard core repulsive interparticle potential, such as either the Yukawa potential (Fig.~\ref{Fig:wvsI_strong}(b)) or the WCA potential from~\cite{Suri2019}, this linear connection holds within an even broader range of $T_{\rm A}$. In particular, we show in Fig.~\ref{Fig:wvsI_strong}(a) that the relation between dissipation and structure for WCA particles persists even close to the beginning of MIPS at a high P\'eclet number, highlighting the robustness of this relationship. (In systems at densities too low to support phase separation, the relation holds across all values of $T_{\rm A}$ that we simulated.)

A natural line of inquiry is to determine whether dissipation can be predicted from pair correlations without polarization or orientation information only for AOUPs and closely related minimal isotropic active matter models or if these results are robust and can be extended to other active systems. We have also performed simulations of actively driven rotors, similar to those in~\cite{vanZuiden2016}, which are anisotropic and have a qualitatively different driving force than AOUPs. Our results suggest that pair correlations provide enough information to predict dissipation for rotors, although unsurprisingly, the usefulness of an observable exactly analogous to $\tilde{I}$ is more limited, especially for short rotors. We leave a more thorough discussion of dissipation and structure in active rotors to future work.

\subsection{Inferring dissipation from static configurations: Insights from neural networks}
\label{neuralnet}

The results in Figs.~\ref{Fig:wvsI_harmonic}-\ref{Fig:wvsI_strong} demonstrate that in certain parameter regimes, the dissipation rate can be inferred by analyzing \emph{solely} the two-point density correlations which characterize static configurations. A natural question which follows is whether it would be possible to train a machine learning algorithm to ``learn'' the dissipation rate from snapshots of active matter.

The field of machine learning applications in chemistry, physics and materials science is a relatively new but flourishing one~\cite{Zdeborova2019, Butler2018, Lee2019}, and much work has been done recently towards using machine learning in active matter~\cite{Cichos2020}. Specifically, a variety of tools have been used to infer patterns, phase boundaries or equations of motions from images or movies of nonequilibrium systems~\cite{Casert2019, Zhao2020, Chase2020, Gnesotto2020, Dulaney2020}. Convolutional neural networks have also been recently used in~\cite{Seif2020} to show how the dissipation rate is connected to time-reversal symmetry violations.

Inspired by these recent successes, we construct a neural network that can map snapshots of particle positions to the underlying dissipation induced by their dynamics. Our goal is to show that the dissipation of generic active systems can be successfully inferred using only these \emph{static} configurations, with no underlying information about the details of the dynamics. Such a proof of principle may then form the basis for the development of feedback control processes, where the dissipation rate is tuned adaptively until the desired structure is achieved.

\begin{figure*}
\centering
    \includegraphics[scale=0.275, clip=True]{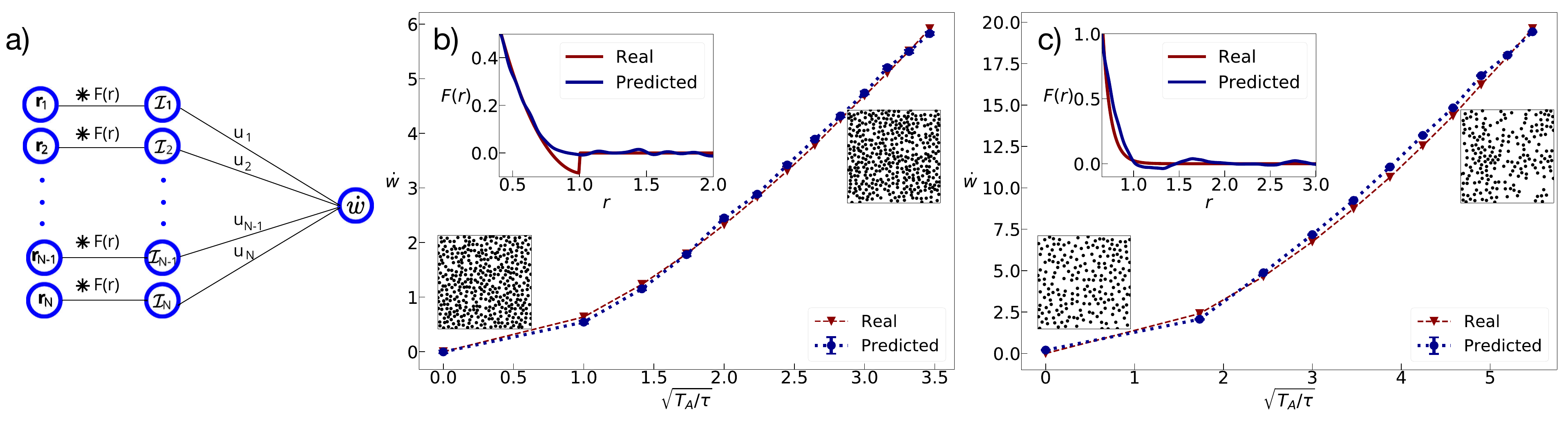}
    \caption{
	Inferring dissipation from static configurations.
	(a)~The machine learning architecture consists of two layers. The first layer is a convolutional layer, which performs the operation ${\bf r}_i * F \equiv \sum_{j \neq i} F(|{\bf r}_i-{\bf r}_j|)$, where $j$ runs over a set number of nearest neighbors. The function $F$ is expressed in a basis of Gaussian functions with coefficients learned by the algorithm (see Materials and Methods). The second layer then fully connects the convolutional layer $\mathcal{I}_i\equiv {\bf r}_i * F$ to the rate of work $\dot w$ through the weight $u$ and bias $b$: $u \sum_i {\mathcal{I}_i} +b = \dot{w} {\rm '}$, where $\dot{w} {\rm '}$ is the scaled rate of work used as the output of the network.
	(b-c)~The machine is trained using AOUP configurations generated at values of $T_{\rm A}\in[0,12]$ (b) or $T_{\rm A}\in[0,30]$ (c), and the values of $\dot w$ are predicted at both the training values of $T_{\rm A}$ and at intermediate values within this range. The agreement between predicted and exact values of $\dot w$ is excellent. This demonstrates that the machine learning algorithm is able to learn dissipation from static snapshots, without any prior information about the underlying dynamics, for AOUPs.
	(b-c inset plots)~We compare the learned function $F$ scaled by an overall deterministic factor $\gamma$, as described in Materials and Methods, with $(\nabla U)^2 - T\nabla^2 U$, showing that the algorithm reproduces the dissipation-structure relation given in~\eqref{eq:dotW_tildeI} without any prior knowledge of such a relationship.
	Parameters for AOUPs: (b) Harmonic potential, $\rho_0=1$, $A=8T$, $\tau=1$, $T=1$, $\gamma=1$; (c)~Yukawa potential, $\rho_0=0.5$,  $A=50T$, $\tau=1$, $\kappa=4$, $T=1$, $\gamma=1$. The other insets show representative snapshots at $T_\A=\{0,10\}$ (Harmonic), and $T_\A=\{0,30\}$ (Yukawa).
	}
    \label{Fig:ML}
\end{figure*}

To this end, we design a network consisting of a continuous convolutional layer followed by a fully connected layer, as outlined in Fig.~\ref{Fig:ML}(a), that maps the input data given by position vectors $\{{\bf r}_i\}$ to the output data given by the scaled rate of work $\dot{w}'$, which is related to the physical work $\dot{w}$ through a simple deterministic transformation (see Materials and Methods). We perform this transformation in order to enhance accuracy and speed of learning. Due to particle indistinguishibility, we constrain the weights of the fully connected layer to be equal. The continuous convolutional layer is inspired by a recent work on rotationally and translationally invariant networks for atomic systems~\cite{tess2018}. This layer effectively scans the neighbors of each particle and sums up the result of applying a learnable function $F({\bf r})$ to each of a number of nearby neighbors, where ${\bf r}$ denotes here the position vector from the particle of interest to its neighbor (see Materials and Methods). This operation is similar to a usual convolution on a grid, with the particle of interest being analogous to the convolution center, and the function $F$ to a filter.

Inspired by the relationship between dissipation and pair correlation functions in~\eqref{eq:dotW_tildeI}, we constrain the form of $F$ to be a function of the distance between particles $r=|{\bf r}|$. Recognizing that the relation between rate of work $\dot w$ and density correlations $h$ given in~\eqref{eq:dotW_tildeI} is also analogous to a convolution, the final learned function $F$, scaled by an overall deterministic factor $\gamma$ (see Materials and Methods), can be expected to be equal to $(\nabla U)^2 - T\nabla^2 U$.

In Figs.~\ref{Fig:ML}(a-b), we show the agreement between the values of $\dot{w}$ measured using the particle-based definition in~\eqref{eq:dotW_microscopic}, referred to as ``exact'' values, and that predicted from static configurations of the system using our neural network. We consider separately configurations taken from either harmonic or Yukawa AOUP particles over a finite range of $T_{\rm A}$. Each data point is averaged over 10 independent machine learning training cycles. The neural network is able to infer the rate of work with good accuracy, which confirms that static configurational data is indeed sufficient to evaluate dissipation. Note that, in the prediction phase, the neural network is supplied with configurations generated at values of $T_{\rm A}$ that are never used in the training phase.

We further demonstrate in Figs.~\ref{Fig:ML}(a-b) (inset plots) how our continuous convolution neural network is able to learn a function $F$ that is very close to the function $(\nabla U)^2 - T\nabla^2 U$. Our simple network, which is designed with the minimum number of elements needed to capture a relationship reminiscent of~\eqref{eq:dotW_tildeI},  succeeds in learning that same relationship with great accuracy. We stress again that our approach does not require any information regarding the microscopic polarization, and it also does not rely on measuring any current, at variance with~\cite{Barato2015, Gingrich2016, Gladrow2016, Li2019}.



\section{Discussion}

Our results demonstrate that the dissipation induced by driving forces in active liquids can be inferred generically by analyzing the static structure of the system. In particular, we analyze the underlying connection between dissipation and two-point density correlations by demonstrating that it exists in a pseudo-equilibrium regime in Section~\ref{smalltau} and extending a mean-field theory for quantitatively predicting the latter to predicting the former in Section~\ref{meanfield}. We then show that the mean-field theory yields predictions matching numerical results specifically for AOUPs with repulsive harmonic interaction potentials, then provide additional numerical results and results from machine learning which suggest the connection is robust across many other regimes.

It would be interesting to explore whether such a relation extends to other types of active liquids, such as for instance liquids with aligning interactions among the particles~\cite{Chate2020}, or with driving forces that sustain a permanent spinning of particles~\cite{vanZuiden2016, Vitelli2018}. We have previously observed that a similar dissipation-structure relation holds when the system is driven by external deterministic forces~\cite{Suri2019}. This suggests that dissipation can be potentially inferred from structure for a large class of nonequilibrium liquids, even beyond active matter.

Importantly, we show that neural networks are able to accurately infer the dissipation by learning only from snapshots of positions. This result illustrates the promise that neural networks, when suitably trained, can capture generic connections between microscopic features and macroscopic variables~\cite{Zdeborova2019, Butler2018}. It also solves the outstanding problem of how to reliably evaluate dissipation based on time-independent and easily accessible data. Indeed, the local polarization of active particles is difficult to access in experimental settings, in particular for isotropic particles, so that obtaining an accurate particle-based measurement of dissipation is generally challenging~\cite{Toyabe2010, Ahmed2016, Mizuno2018}. Besides, measuring currents to deduce bounds on dissipation, as proposed in~\cite{Barato2015, Gingrich2016, Gladrow2016, Li2019}, has only limited applications to active systems without any obvious observable current, such as the one considered here.


This work was mainly funded by support from a DOE BES Grant DE-SC0019765 to LT, GR and SV (Theory,simulations and Machine learning). EF was funded by the Luxembourg National Research Fund (FNR), grant reference 14389168.


\section*{Materials and methods}

\subsection*{Numerical simulations}

For sections pertaining to strong interactions and relationship between rate of work and density correlations, the simulations are run in a two-dimensional box $10^2\sigma\times 10^2\sigma$ with periodic boundary conditions, where $\sigma = 1$ is the particle diameter. The time step for harmonic and WCA simulations is $\delta t = 10^{-4}$, and that for Yukawa simulations is $\delta t = 10^{-6}$; in all cases, the initial condition is homogeneous. The density is $\rho_0 = 0.5$ when the Yukawa potential is used, 0.75 when WCA is used, and 1 when the harmonic potential is used. The pair correlation function was extracted over a range of $r \in [0, 3.2]$  in increments of $dr = 0.01$ for WCA particles, and $r \in [0, L/2]$ for Yukawa and harmonic particles, where $L$ denote here the system size. Between 5 and 25 trials are conducted for each set of parameters.

Whenever the harmonic potential is used, the equations of motion are integrated using a custom code of molecular dynamics, based on finite time difference. The harmonic systems are equilibrated for 500 units of simulation time, corresponding to at least $500 \tau$ for all simulations, where $\tau$ is the persistence time of the active noise, and data is collected every 100 units for a duration of 1000 time steps.

Whenever the Yukawa or WCA potentials are used, the simulations are performed using the LAMMPS simulation package with a custom fix to allow for self-propulsion and custom code to measure pair correlation functions and dissipation in separate simulations. In the case of WCA, equilibration is performed for at least $50\tau$, pair correlation functions are measured for at least $100\tau$, and dissipation is measured for at least $25\tau$. In the case of Yukawa, equilibration is performed for $12\tau$, pair correlation functions are measured for at least $24\tau$, and dissipation is measured for $6\tau$.

Calculation of $\tilde{I}$ from simulations is performed by numerically integrating the difference between the $g(r)$ and $g_{\rm eq}(r)$ histograms from simulations, multiplied by derivatives of the interparticle potential following~\eqref{eq:dotW_tildeI}. These $g(r)$ are obtained by generating histograms of distances between each pair of particles with resolution $dr = 0.01\sigma$, averaged over at least 5 independent trials with at least 11 snapshots per trial over the domains mentioned above. Calculation of $c_{\rm eq}(k)$ as input for theoretical predictions is done by numerically Fourier transforming the portion of the equilibrium $h_{\rm eq}(r) = g_{\rm eq}(r) - 1$ with $r \in [0, 16]$ to obtain $h_{\rm eq}(k)$, and then computing $c_{\rm eq}(k)$ using the Ornstein-Zernike relation (shown in~\eqref{eq:c} extended to non-equilibrium systems). Equilibrium $g(r)$ for these calculations are obtained by generating histograms of distances between each pair of particles with resolution $dr = 0.01\sigma$, averaged over 15 independent trials with 11 snapshots per trial and limited to $r \in [0, 20]$; the simulation box has size $40\sigma\times 40\sigma$. Fourier transformation to obtain $h_{\rm eq}(k)$ is done by multiplying $h_{\rm eq}(r)$ by $2 \pi r J_0(k \cdot r)$ and integrating over $r \in [0, 16]$, repeated for $k \in [2\pi/16, 16\pi]$ incremented by $2\pi/16$.

The simulations used to extract snapshots for machine learning were run in a two-dimensional box of size $25\sigma\times 25\sigma$ with periodic boundary conditions, using the custom code referenced previously. The systems were equilibrated for $500 \tau$ and snapshots were saved every $\tau$ for a total simulation time of $20 \tau$. Multiple trajectories were generated so as to obtain the 6000 snapshots per each $T_{\rm A}$ required for training and testing.


\subsection*{Continuous convolutional neural network}

The network was built in keras using the Functional API~\cite{keras}. The individual input to the network consists of interparticle distances. This choice was made because it greatly simplified the implementation of a custom layer in keras. To be more precise, our input consists of an array of size $N \times D$ that contains, for each of the $N$ particles, the distances to its nearest $D$ neighbors. For the simulations with the harmonic potential, $N = 625$ and $D = 20$, while for the simulations with the Yukawa potential, $N = 312$ and $D = 25$. The choice of $D$ was such that distances to around $r = 3$ ($r = 4$ for Yukawa) are mostly captured, but with certainty all distances up to the cutoff ($r = 1$ for harmonic and $r=2.5$ for Yukawa) are captured. This choice ensures that the algorithm has access to all the distances that we expect are crucial to extracting the rate of work.

The continuous convolutional layer is built as a custom layer and performs the following. For each convolution center, in this case the position of active particle ${\bf r}_i$, the convolution operation consists of evaluating the sum $\sum_{j=1}^{D} F(|{\bf r}_i - {\bf r}_j|)$. Here, $F$ is a learnable function, the equivalent of a filter in traditional convolutional neural networks. We express this function in a basis of $N_G=30$ ($N_G=40$  for Yukawa) Gaussian functions centered between $r=0$ and $r=3$ ($r = 4$ for Yukawa) and with a standard deviation of 0.05:
\begin{equation}\label{eq:Fr}
    F(r) = \sum_{i=0}^{N_G-1} \beta_i e^{-(r-0.1 i)^2/(2\cdot 0.05^2)} .
\end{equation}
Finally, the output to the network consists of the rate of work $\dot{w}$, centered as
\begin{equation}\label{eq:dw_recenter}
    \dot{w}_{\text{c}} = \dot{w} + \alpha \rho_0 \int \big[(\nabla U)^2 - T \nabla^2 U \big] g_{\rm{eq}} d{\bf r} ,
\end{equation}
and scaled as
\begin{equation}\label{eq:dw_rescale}
    \dot{w} {\rm '} = \dfrac{\dot{w}_{\text{c}}}{\max(\dot{w}_{\text{c}}) - \min(\dot{w}_{\text{c}} )},
\end{equation}
where the maximum and minimum of the rate of work is taken over the entire training data set. Centering and scaling the rate of work improves the accuracy and speed of our network. The rate of work is calculated at each value of $T_{\rm{A}}$ as an average in the nonequilibrium steady state of~\eqref{eq:dotW_microscopic}. Hence, different snapshots at the same value of $T_{\rm{A}}$ will be mapped to the same output.

In conclusion, the machine performs the following mapping from the input (interparticle distances $r_{ij} = |{\bf r}_i-{\bf r}_j|$) to the output (scaled rate of work):
\begin{equation}\label{eq:final_rel}
    u \sum_{i,j} \sum_{k=0}^{N_G-1} \beta_k e^{-(r_{ij}-0.1 k)^2/(2\cdot 0.05^2)} + b = \dot{w} {\rm '} .
\end{equation}
We enforce particle indistinguishibility by setting a layer constraint that the weights $u$ are identical. The machine is tasked with finding the best $\{u, \beta_k, b\}$ to minimize the deviation between its predicted output, $\dot{w}_{\text{pred}}$, and $\dot{w} {\rm '}$. We choose to quantify this deviation through a loss function of the form
\begin{equation}
    L(\{\dot{w}^a_{{\rm pred}, i}\}) = \sum_a \left[  \left( \dfrac{\sum_i \dot{w}^a_{{\rm pred}, i}}{N_a} \right) - \dot{w}'_a \right]^2,
\end{equation}
where the $a$ indexes the $T_{\rm{A}}$ values used in the training process, and $N_a$ is the number of snapshots at that value of $T_{\rm{A}}$ in each training batch over which the loss function is calculated. Our choice of loss function ensures that we are mapping the average of the output of the  network at each $T_{\rm A}$ to the rate of work at the same $T_{\rm A}$. Since the rate of work is indeed an average function of the configurations, this choice of loss function was the most suitable for training our model. In the large $N$ limit, we have the following relation: 
\begin{equation}
    u \sum_{i,j} F(r_{ij}) = Nu \rho_0 \int F(r) g({\bf r}) d{\bf r} .
\end{equation}
If the training is successful and the machine predicts outputs that are very close to $\dot{w} {\rm '}$, then using~\eqref{eq:dotW_tildeI} along with~(\ref{eq:Fr}-\ref{eq:final_rel}), we recover:
\begin{equation}
    \begin{aligned}
        N u &\rho_0 \int F g d{\bf r} + b =
        \\
        & \frac{\alpha \rho_0}{\max(\dot{w}_{\text{c}}) - \min(\dot{w}_{\text{c}})} \int \big[(\nabla U)^2 - T \nabla^2 U\big]  g d{\bf r} ,
    \end{aligned}
\end{equation}
yielding
\begin{equation}\label{eq:param_ML}
    \begin{aligned}
    	b &= 0,
    	\quad
    	\gamma_{NN} F = (\nabla U)^2 - T\nabla^2 U ,
    	 \\
    	\gamma_{NN} &= \frac{N u}{\alpha} \big[ \max(\dot{w}_{\text{c}}) - \min(\dot{w}_{\text{c}}) \big] .
	\end{aligned}
\end{equation}
The good agreement between $\gamma_{NN} F$ and $(\nabla U)^2 - T\nabla^2 U$ shown in Fig.~\ref{Fig:ML} confirms the relations in~\eqref{eq:param_ML}.

In producing the data of Fig.~\ref{Fig:ML}(a-b), we prepared 35000 snapshots using the harmonic potential and 30000 snapshots using the Yukawa potential. These correspond to 5000 snapshots per $T_{\rm A}$, where $T_{\rm A}$ ranged from 0 to 12 in increments of 2 for harmonic and from 0 to 30 in increments of 6 for Yukawa. The snapshots were divided into 80\% train and 20\% validation sets, and the network was trained until we observed convergence of the validation loss (between 200 and 300 epochs of training), using the adam optimizer in keras with batch size 512.

The test data reflected in Figs.~\ref{Fig:ML}(b-c) comes from independent simulations ran over a more fine-grained range of $T_{\rm A}$'s. Specifically, for harmonic we extracted 1000 configurations per $T_{\rm A}$ with $T_{\rm A}$ ranging over the integers between 0 and 12. For Yukawa we extracted 1000 configurations per $T_{\rm A}$ with $T_{\rm A}$ going between 0 and 30 in increments of 3.

\subsection*{Code}

Codes for molecular dynamics and machine learning can be found at \url{https://github.com/ltociu/structure_dissipation_active_matter}.

\bibliography{References}

\appendix
\section{Gaussian density fluctuations}
\label{sec:appA}

Density correlations $\rho$ can be expressed in terms of their mean value $\langle\rho\rangle = \rho_0$ and fluctuations $\delta \rho=\rho-\rho_0$. To obtain an expression for $\tilde{I}$ that is directly related to $\dot{w}$ and depends only on pair correlations, we assume that $\delta \rho$ is small and Gaussian in nature. This is most justifiable when $\rho_0$ is large and interactions between particles are relatively weak. Note that in the limit of small $\rho_0$, contributions from three-body terms in~\eqref{eq:dotW_tildeI_Gaussian} are suppressed and the relation between $\dot{w}$ and $\tilde{I}$ is simple.

Applying our approximation to three-body density correlations, we arrive at
\begin{equation}
\begin{aligned}
    &\rho_0^3 g_3\left(\mathbf{r},\mathbf{r'},\mathbf{r''}\right)
    \\
    &= \langle \rho\left(\mathbf{r}\right) \rho\left(\mathbf{r'}\right) \rho\left(\mathbf{r''}\right) \rangle
     \\
    &= \rho_0^3 + \rho_0^2 \langle \delta\rho\left(\mathbf{r}\right) + \delta\rho\left(\mathbf{r'}\right) + \delta\rho\left(\mathbf{r''}\right) \rangle \\
    & \quad + \rho_0 \langle \delta\rho\left(\mathbf{r}\right)\delta\rho\left(\mathbf{r'}\right) + \delta\rho\left(\mathbf{r'}\right)\delta\rho\left(\mathbf{r''}\right) + \delta\rho\left(\mathbf{r''}\right)\delta\rho\left(\mathbf{r}\right) \rangle\\
    & \quad+ \langle \delta\rho\left(\mathbf{r}\right)\delta\rho\left(\mathbf{r'}\right)\delta\rho\left(\mathbf{r''}\right) \rangle \\
    & \approx \rho_0^3 + \rho_0 \langle \delta\rho\left(\mathbf{r}\right)\delta\rho\left(\mathbf{r'}\right) + \delta\rho\left(\mathbf{r'}\right)\delta\rho\left(\mathbf{r''}\right) + \delta\rho\left(\mathbf{r''}\right)\delta\rho\left(\mathbf{r}\right) \rangle ,
\end{aligned}
\end{equation}
in which we have assumed that the term with $\delta\rho^3$ is negligibly small and used that $\langle\delta\rho\rangle=0$. We can similarly approximate two-body density correlations to arrive at
\begin{equation}
\begin{aligned}
    \rho_0^2 g\left(\mathbf{r},\mathbf{r'}\right) &= \langle \rho\left(\mathbf{r}\right) \rho\left(\mathbf{r'}\right) \rangle
    \\
    &= \rho_0^2 + \rho_0 \langle \delta\rho\left(\mathbf{r}\right) + \delta\rho\left(\mathbf{r'}\right) \rangle + \langle \delta\rho\left(\mathbf{r}\right)\delta\rho\left(\mathbf{r'}\right) \rangle \\
    & = \rho_0^2 + \langle \delta\rho\left(\mathbf{r}\right)\delta\rho\left(\mathbf{r'}\right) \rangle .
\end{aligned}
\end{equation}
Combining these results, we get
\begin{equation}
    g_3\left(\mathbf{r},\mathbf{r'},\mathbf{r''}\right) \approx g\left(\mathbf{r},\mathbf{r'}\right) + g\left(\mathbf{r'},\mathbf{r''}\right) + g\left(\mathbf{r''},\mathbf{r}\right) - 2 ,
\end{equation}
which in turn means that, after changing function arguments to be consistent with definitions of $g$ and $g_3$ in~\eqref{eq:g2g3},
\begin{equation}
    \Delta g_3\left(\mathbf{r},\mathbf{r'}\right) \approx \Delta g\left(\mathbf{r}-\mathbf{r'}\right) + \Delta g\left(\mathbf{r}\right) + \Delta g\left(\mathbf{r'}\right) .
\end{equation}
When we insert this expression for $\Delta g_3$ into~\eqref{eq:dotW_tildeI_exact}, the $g\left(\mathbf{r}\right)$ and $g\left(\mathbf{r'}\right)$ terms yield 0 when multiplied by $\nabla U\left(\mathbf{r'}\right)$ and $\nabla U\left(\mathbf{r}\right)$ and integrated over $\mathbf{r'}$ and $\mathbf{r}$ respectively. Only the $\Delta g\left(\mathbf{r}-\mathbf{r'}\right)$ term yields a non-zero contribution to $\dot{w}-\tilde{I}$, and thus we obtain~\eqref{eq:dotW_tildeI_Gaussian} as our final expression from the assumption of Gaussian density fluctuations.

\renewcommand{\thefigure}{A\arabic{figure}}
\setcounter{figure}{0}
\begin{figure*}
    \centering
    \includegraphics[scale=0.096, clip=True]{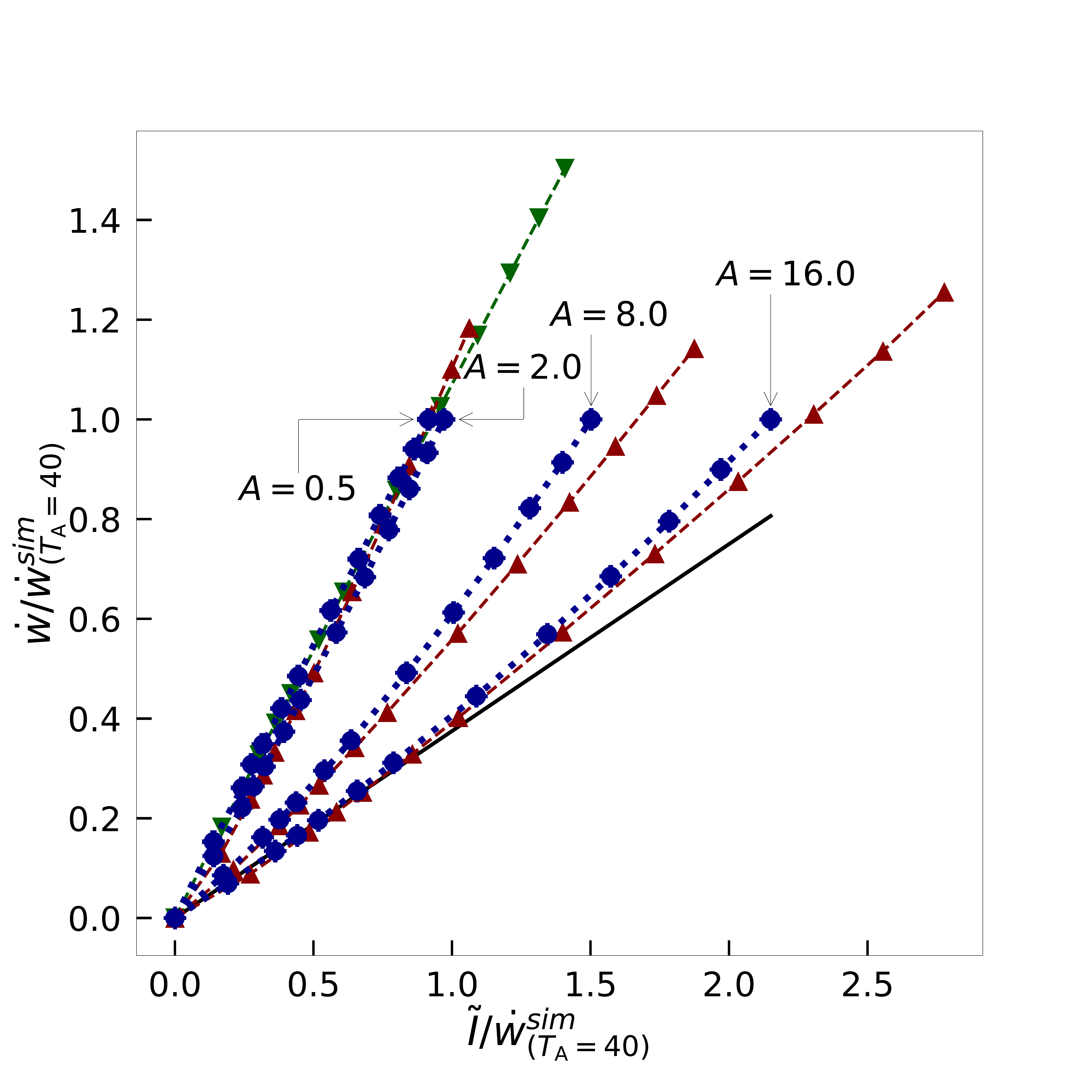}
    \caption{
	Simulation results for $\dot{w}$ versus $\tilde I$ for particles with the harmonic interparticle potential at various amplitudes (labeled) scaled by the value of $\dot{w}$ at $T_{\rm{A}} = 40$. All details other than the range of $T_{\rm A}$ and the scaling factor are identical to Fig.~\ref{Fig:wvsI_harmonic}. In addition, the solid black line indicates what a true linear relationship between $\dot{w}$ and $\tilde{I}$ for $A=16$ would be if their constant of proportionality at $T_{\rm{A}}=5$ held across all values of $T_{\rm{A}}$. Deviations of both theoretical and simulation results from this line illustrate the breakdown of the linear relationship as $T_{\rm{A}}$ increases. Our theory still captures this nonlinear relationship between $\dot{w}$ and $\tilde{I}$ reasonably well, but deviations are more significant than when considering $T_{\rm{A}} \leq 10$. Parameters: harmonic potential, $\rho_0=1$, $\tau=1$, $T=1$, $\gamma=1$ $T_{\rm{A}}\in[0,40]$.
    }
    \label{Fig:A1}
\end{figure*}

\section{Mean-field theory for dissipation}
\label{sec:appB}
\subsection{Density field for weakly interacting tracer}

This derivation reproduces large parts of same conceptual and mathematical framework we described in our companion paper~\cite{Tociu2022} and merely extends them to obtain expressions of $\dot{w}$ in addition to $h(\mathbf{k})$. We start by considering the effective dynamics of an active tracer weakly interacting with a bath consisting of the other particles. To analytically derive the statistics of the tracer displacement, our strategy, inspired by recent works~\cite{Demery2011, Demery2014}, is to scale the interaction strength by a dimensionless factor $\varepsilon$, which can be regarded as a small parameter for perturbative expansion. The equation of motion of the tracer position ${\bf r}_0(t)$ then reads
\begin{equation}\label{eq:dyn_pert}
    \dot{\bf r}_0 = \frac{{\bf f}_0}{\gamma} - \frac{\varepsilon}{\gamma} \int \nabla_0 \U({\bf r}_0  - {\bf r}') \rho ({\bf r}', t)  d {\bf r}' + {\bm \xi}_0 ,
\end{equation}
where the bath is described in terms of the density field $\rho ( {\bf r} , t) = \sum_{i=1}^N \delta({\bf r} - {\bf r}_i(t))$ where $N$ is the number of bath particles. The dynamics of the density field $\rho({\bf r} , t)$, can be readily obtained following the procedure in~\cite{Dean_1996} as
\begin{align*}\label{eq:field_EOM}
    &\gamma\dfrac{\partial \rho ({\bf r}, t)}{ \partial t} = \T \nabla^2 \rho({\bf r}, t) + \nabla \cdot \big[ \sqrt{2 \rho_0 \T \gamma} {\bm \Lambda}({\bf r}, t) -\bP({\bf r}, t) \big] 
    \\
    &\quad + \nabla \cdot \left( \rho \nabla \left[ \int  \U({\bf r}  - {\bf r}') \rho({\bf r}', t)  d {\bf r}' + \varepsilon\U({\bf r} - {\bf r}_0)  \right] \right) , \numberthis
\end{align*}
where $\bP$ denotes polarization field $ \bP ({\bf r}, t) = \sum_i {\bf f}_i(t) \delta({\bf r} - {\bf r}_i(t))$ and $\boldsymbol\Lambda$ is a Gaussian white noise with zero mean and unit variance ($\langle \Lambda_{\alpha} ({\bf r}, t) \Lambda_{\beta} ({\bf r}', t') \rangle = \delta_{\alpha \beta} \delta({\bf r} - {\bf r}') \delta(t - t')$). In principle, the dynamics~(\ref{eq:dyn_pert}-\ref{eq:field_EOM}) can be solved recursively to obtain the statistics of the density field $\rho({\bf r} , t)$ and of the tracer position ${\bf r}_0$. Some of us already took this approach in~\cite{Suri2019, Suri2020} using a perturbation in the weak interaction limit. In what follows, we extend this approach to characterize the system beyond the regime of weak interactions.


\subsection{Derivation of dissipation}

Our nonequilibrium mean-field theory to solve for $\dot{w}$ is built as follows. Neglecting polarization, we first linearize the dynamics~\eqref{eq:dyn_pert}-\eqref{eq:field_EOM} and obtain a solution for $\delta \rho$ in the Fourier domain. We then obtain a formula for $\dot{w}$ in terms of $\delta \rho$ and construct an expansion in the coupling parameter $\varepsilon$ to compute $\dot{w}$ up to second order in $\varepsilon$. Finally, we draw inspiration from equilibrium solvation theories~\cite{Chandler1993} to extend this theory from weakly interacting to strongly interacting particles using an approximate method numerically validated in~\cite{Tociu2022}.

We begin with~\eqref{eq:field_EOM} and do not consider the polarization term further. Our choice is justified in the low-activity limit and, beyond that, supported by the theoretical and numerical results in~\cite{Suri2020} for efficiency and mobility even with strong driving forces present. Linearizing the dynamics of the density field $\rho$ around the overall density $\rho_0$, we arrive at closed-form equation of motion for $ \delta\rho = \rho - \rho_0$. This linear approximation holds for weak interparticle potentials, so that any local density fluctuation is small compared to the density of the liquid. The solution for $\delta \rho({\bf k}, t) = \int [ \rho({\bf r}, t) - \rho_0 ] e^{-i{\bf k} \cdot {\bf r}} d{\bf r}$ follows readily as:
\begin{equation}\label{eq:red_field_EOM_k}
\begin{aligned}
    &\gamma \delta \rho( {\bf k}, t) = \int_{-\infty}^t ds e^{-{\bf k}^2 G({\bf k}) (t-s)} \\
    &\qquad  \times\left( -{\bf k}^2 \rho_0 \varepsilon \U({\bf k}) e^{-i{\bf k} \cdot {\bf r}_0(s)} + i{\bf k} \cdot \sqrt{2\rho_0 \T \gamma} \bm{\Lambda} ({\bf k}, s) \right) ,
\end{aligned}
\end{equation}
where $G({\bf k}) =  T + \rho_0 U({\bf k})$, and $\boldsymbol\Lambda$ is a zero-mean Gaussian white noise with correlations
\begin{equation}\label{eq:field_noise_k}
    \langle \Lambda_{\alpha}({\bf k}, s)  \Lambda_{\beta}({\bf k}', s')  \rangle = (2\pi)^d \delta_{\alpha \beta}\delta(s-s')\delta({\bf k} + {\bf k}') ,
\end{equation}
where $d$ is the spatial dimension.

\begin{widetext}
Substituting into the dissipation per particle~\eqref{eq:dotW_microscopic} with the coupling parameter $\varepsilon$ included,
\begin{equation*}
    \dot{w} = \frac{1}{N\gamma} \sum_{j \neq i} \langle {\bf f}_i \cdot \nabla_i \varepsilon U({\bf r}_i-{\bf r}_j) \rangle = -  \int \frac{d{\bf k}}{\gamma (2\pi)^d} \Big\langle {\bf f}_0(t) \cdot i{\bf k} \varepsilon \U({\bf k}) \,\delta \rho({\bf k}, t) e^{ i {\bf k}\cdot{\bf r}_0(t) } \Big\rangle,
\end{equation*}
yields the following:
\begin{equation}\label{eq:dotW_MF}
\begin{aligned}
    \dot{w} &=  \frac{\rho_0}{\gamma} \int \frac{d{\bf k}}{(2\pi)^d} {\bf k}^2 (\varepsilon U({\bf k}))^2 \int_{-\infty}^0 ds e^{{\bf k}^2 G({\bf k})s} \Big\langle i{\bf k} \cdot {\bf f}_0(0) e^{i{\bf k}\cdot ({\bf r}_0(0) - {\bf r}_0(s))}\Big\rangle \\
    & \quad + \sqrt{2\rho_0 \T \gamma} \int \frac{d{\bf k}}{(2\pi)^d} {\bf k}^2  \varepsilon U({\bf k}) \int_{-\infty}^0 ds e^{{\bf k}^2 G({\bf k})s} \left \langle e^{i{\bf k}\cdot {\bf r}_0(0)} {\bf f}_0(0) \cdot  {\bm \Lambda}({\bf k}, s)\right \rangle .
\end{aligned}
\end{equation}
Next, we solve for the tracer position ${\bf r}_0$ to yield an expression which depends instead on driving forces ${\bf f}_0$, thermal noise ${\bm \xi}_0$, and interparticle potential $\U({\bf k})$. From the tracer dynamics~\eqref{eq:dyn_pert}, we deduce
\begin{equation}\label{eq:tracer}
    {\bf r}_0(0) = \int_{-\infty}^0 [ {\bf f}_0(x) + {\bm \xi}_0(x)] dx + \varepsilon \int \frac{d{\bf k}'}{(2\pi)^d} i{\bf k}' \U({\bf k}') \int_{-\infty}^0 ds' \,\delta \rho({\bf k}', s') e^{ i {\bf k}\cdot{\bf r}_0(s')} ,
\end{equation}
from which, after expanding with respect to the parameter $\varepsilon$, we derive
\begin{equation}\label{eq:tracer_bis}
    e^{i{\bf k}\cdot{\bf r}_0(0)} = e^{i{\bf k}\cdot \int_{-\infty}^0 [ {\bf f}_0(x) + {\bm \xi}_0(x)] dx}  \bigg[ 1 - \varepsilon \int \frac{d{\bf k}'}{(2\pi)^d} {\bf k}\cdot{\bf k}' \U({\bf k}') \int_{-\infty}^0 ds' \delta \rho({\bf k}', s') e^{i{\bf k}'\cdot \int_{-\infty}^{s'} [ {\bf f}_0(x) + {\bm \xi}_0(x)] dx} + {\cal O}(\varepsilon^2) \bigg] .
\end{equation}
Substituting in the expression for $\delta\rho$ given in Eq.~\eqref{eq:red_field_EOM_k}, we obtain
\begin{equation}\label{eq:tracer_ter}
\begin{aligned}
    e^{i{\bf k}\cdot{\bf r}_0(0)} = & e^{i{\bf k}\cdot \int_{-\infty}^0 [ {\bf f}_0(x) + {\bm \xi}_0(x)] dx} \bigg[ 1 - \varepsilon \int \frac{d{\bf k}'}{(2\pi)^d} ({\bf k}\cdot{\bf k}') \U({\bf k}') \int_{-\infty}^0 ds' e^{i{\bf k}'\cdot \int_{-\infty}^{s'} [ {\bf f}_0(x) + {\bm \xi}_0(x)] dx} \times \\
    & \int_{-\infty}^{s'} ds'' e^{-{\bf k}'^2 G({\bf k}') (s'-s'')} i {\bf k'} \cdot \sqrt{2T\rho_0} {\bm\Lambda}({\bf k}',s'') + {\cal O}(\varepsilon^2) \bigg] .
\end{aligned}
\end{equation}
Substituting~\eqref{eq:tracer_ter} into~\eqref{eq:dotW_MF}, truncating at second order in $\varepsilon$, and letting $\gamma=1$ for simplicity, we then get
\begin{align}\label{eq:dotW_MF_full}
    \dot{w} & =  \rho_0  \int \frac{d{\bf k}}{(2\pi)^d} {\bf k}^2 ( \varepsilon U({\bf k}))^2 \int_{-\infty}^0 ds e^{{\bf k}^2 G({\bf k})s}  \Big\langle i{\bf k} \cdot {\bf f}_0(0)   e^{i{\bf k}\cdot \int_s^0 [ {\bf f}_0(x) + {\bm \xi}_0(x)] dx}  \Big\rangle \nonumber \\
    & - {2\rho_0 \T} \int \frac{d{\bf k}d{\bf k}'}{(2\pi)^{2d}} {\bf k}^2 ({\bf k}\cdot{\bf k}') \varepsilon^2 U({\bf k}) U({\bf k}') \int_{-\infty}^0 ds' e^{ -{\bf k'}^2 G({\bf k}') s'} \Big\langle  i{\bf k}' \cdot {\bf f}_0(0) e^{i{\bf k}\cdot \int_{-\infty}^0 [ {\bf f}_0(x) + {\bm \xi}_0(x) ]dx+i{\bf k}'\cdot \int_{-\infty}^{s'} [{\bf f}_0(x) + {\bm \xi}_0(x)] dx}  \Big\rangle \nonumber \\
    & \times \int_{-\infty}^{0} ds  \int_{-\infty}^{s'} ds'' e^{ {\bf k}^2 G({\bf k}) s + {\bf k}'^2 G({\bf k}') s''}  \langle \Lambda_\alpha({\bf k}, s) \Lambda_\alpha({\bf k}', s'')\rangle ,
\end{align}
where we have used that $\bm\Lambda$, ${\bm\xi}_0$, and ${\bf f}_0$ are independent.
\end{widetext}
We simplify this by observing that according to Wick's theorem, we can write for the white noise
\begin{equation}\label{eq:expxi}
    \left \langle e^{i{\bf k}\cdot \int_{s}^0 {\bm \xi}_0(x) dx} \right\rangle =e^{{\bf k}^2 Ts} .
\end{equation}
To treat the equivalent terms for the active forces, we start from the time correlations~\eqref{eq:EOM_noise} and derive the following:
\begin{equation}
    \left \langle \int_{s}^0 f_{0\alpha}(0) f_{0\alpha} (x) dx \right \rangle = T_{\rm A} (1 - e^{s/\tau}) ,
\end{equation}
and
\begin{equation}\label{eq:expxi2}
\begin{aligned}
    &\left \langle \int_{s}^0 \int_{s}^0  f_{0\alpha} (x)  f_{0\alpha} (x')  dx dx' \right \rangle
    \\
    &= \frac{T_{\rm A}}{\tau} \int_{s}^0 dx' \Bigg( \int_{x}^0 e^{-(x'-x)/\tau} dx + \int_{s}^x e^{-(x-x')/\tau} dx \Bigg)
    \\
    &= - 2 \big[ T_{\rm A} s + T_{\rm A} \tau ( 1- e^{s/\tau}) \big] \equiv - 2 R(s) .
\end{aligned}
\end{equation}
Again according to Wick's theorem, we can now write for the driving forces:
\begin{equation}\label{eq:expf}
    \left\langle e^{i{\bf k}\cdot \int_{s}^0 {\bf f}_0(x) dx} \right\rangle =e^{{\bf k}^2 R(s)} ,
\end{equation}
and
\begin{equation}\label{eq:f_time_expfSI}
    \begin{aligned}
    &\left \langle i{\bf k} \cdot {\bf f}_0(0) e^{i{\bf k} \int_s^0 {\bf f}_0(x) dx} \right \rangle
    \\
    &\qquad= -{\bf k}^2 \left \langle  \int_s^0 f_{0\alpha}(0) f_{0\alpha}(x) dx \right \rangle e^{{\bf k}^2R(s)}
    \\
    &\qquad=-{\bf k}^2 e^{{\bf k}^2 R(s)} T_{\rm A} (1-e^{s/\tau}) .
    \end{aligned}
\end{equation}
Taken together, this means that we can make the simplifications of the following kind:
\begin{multline}
    \Big \langle e^{i{\bf k}\cdot \int_{-\infty}^0 [ {\bf f}_0(x) + {\bm \xi}_0(x)] dx - i{\bf k}\cdot \int_{-\infty}^s [ {\bf f}_0(x) + {\bm \xi}_0(x)] dx} \Big \rangle
    = \\ e^{{\bf k}^2 \left(Ts + R(s)\right)} .
\end{multline}
After collapsing the noise correlation functions and using~\eqref{eq:field_noise_k},~\eqref{eq:expxi},~\eqref{eq:expf} in this manner to simplify~\eqref{eq:dotW_MF_full}, we obtain
\begin{equation}
\begin{aligned}
    \dot{w} &= - \rho_0 T_{\rm A} \int \frac{d{\bf k}}{(2\pi)^d} {\bf k}^4 (\varepsilon U({\bf k}))^2 
    \\
    \quad &\times\bigg[ \int_{-\infty}^0 ds e^{{\bf k}^2 (( G({\bf k}) + T)s + R(s))} (1-e^{s/\tau}) + \\
    & \qquad \dfrac{T}{G({\bf k})} \int_{-\infty}^0 ds'  e^{{\bf k}^2 (( G({\bf k}) + T)s' + R(s'))} (1-e^{s'/\tau}) \bigg] ,
\end{aligned}
\end{equation}
where $G({\bf k}) = T + \rho_0 U(({\bf k}) $. We simplify this a bit further and set $\varepsilon$ to 1 to obtain~\eqref{eq:dotW_MF_final}, an expression valid for systems with weak interparticle potentials and high densities:
\begin{align}\label{eq:dotW_MF_final}
    \dot{w} &= - \rho_0 T_{\rm A} \int \frac{d{\bf k}}{(2\pi)^d} {\bf k}^4 (\varepsilon U({\bf k}))^2 \dfrac{ G({\bf k}) + T}{G({\bf k})} \nonumber \\
    & \quad\times\int_{-\infty}^0 ds e^{{\bf k}^2 ( \left(G({\bf k}) + T)s + R(s)\right)}(1-e^{s/\tau}) .
\end{align}
As explained in~\cite{Tociu2022}, a similar procedure is followed to obtain the mean-field expression for $h\left(\mathbf{k}\right)$:
\begin{equation}\label{eq:hk_MF}
    h({\bf k}) = -  {\bf k}^2 \varepsilon U({\bf k})  \dfrac{ G({\bf k}) + T}{G({\bf k})} \int_{-\infty}^0 ds e^{{\bf k}^2 \left( (G({\bf k}) + T)s + R(s)\right)} .
\end{equation}
To go beyond the regime of weak interactions and obtain~\eqref{eq:hk_MF_c} and~\eqref{eq:dotW_MF_c}, we substitute $U(\bf{k})$ where it arises from~\eqref{eq:red_field_EOM_k} with $-T c_{\rm eq}(\bf{k})$. This procedure is inspired by equilibrium solvation theories~\cite{Chandler1993}, in which context the density around a tracer particle interacting \emph{strongly} with its neighbors is captured by considering the convolution between the density correlation and direct correlation functions. In the limit of weak interactions, these functions are approximately equivalent (as mentioned in Section~\ref{smalltau}) and therefore this substitution is safe.
\end{document}